\documentclass[conference]{IEEEtran}
\usepackage{cite}
\usepackage{url}
\usepackage{amsmath,amssymb,amsfonts}
\usepackage{algorithmic}
\usepackage{textcomp}
\usepackage{caption}
\def\BibTeX{{\rm B\kern-.05em{\sc i\kern-.025em b}\kern-.08em
    T\kern-.1667em\lower.7ex\hbox{E}\kern-.125emX}}
    
\usepackage{ifpdf}
 \ifpdf
\usepackage[pdftex]{graphicx}
 \else
\usepackage[dvips]{graphicx}
 \fi
 
\def\bitcoinA{
  \leavevmode
  \vtop{\offinterlineskip
    \setbox0=\hbox{B}
    \setbox2=\hbox to\wd0{\hfil\hskip-.03em
    \vrule height .3ex width .15ex\hskip .08em
    \vrule height .3ex width .15ex\hfil}
    \vbox{\copy2\box0}\box2}}

\begin{document}
    
\title{The Evolution of Embedding Metadata \\ in Blockchain Transactions \\}

\author{
\IEEEauthorblockN{Tooba Faisal}
\IEEEauthorblockA{\textit{University College London} \\
London, UK\\
tooba.hashmi@gmail.com}
\and
\IEEEauthorblockN{Nicolas Courtois}
\IEEEauthorblockA{\textit{University College London} \\
London, UK\\
n.courtois@ucl.ac.uk}
\and
\IEEEauthorblockN{Antoaneta Serguieva}
\IEEEauthorblockA{\textit{nChain, and LSE Systemic Risk} \\
London, UK\\
antoaneta@ncrypt.com}
}

\maketitle
 
\begin{abstract}
The use of blockchains is growing every day, and their utility has greatly expanded from sending and receiving crypto-coins to smart-contracts and decentralized autonomous organizations. Modern blockchains underpin a variety of applications: from designing a global identity to improving satellite connectivity. In our research we look at the ability of blockchains to store metadata in an increasing volume of transactions and with evolving focus of utilization. We further show that basic approaches to improving blockchain privacy also rely on embedding metadata. This paper identifies and classifies real-life blockchain transactions embedding metadata of a number of major protocols running essentially over the bitcoin blockchain. The empirical analysis here presents the evolution of metadata utilization in the recent years, and the discussion suggests steps towards preventing criminal use. Metadata are relevant to any blockchain, and our analysis considers primarily bitcoin as a case study. The paper concludes that simultaneously with both expanding legitimate utilization of embedded metadata and expanding blockchain functionality, the applied research on improving anonymity and security must also attempt to protect against blockchain abuse.
\end{abstract}


\section{Introduction}
The use of blockchains is expanding from transferring crypto-coins to implementing smart-contracts that service a variety of domains. IBM and Sovrin are designing and implementing a global digital identity layer enabled by blockchain: decentralized, point-to-point exchange of information about people, organizations, or things. \cite{ibm} EtherSat is developing a protocol for satellite connectivity utilizing blockchain: a decentralized global area network that maximizes efficiency of existing ground-station infrastructure. \cite{sat} nChain is creating a blockchain tokenization layer to enable interactivity and interoperability among smart contracts underlying various services \cite{chain1}\cite{chain2}. These are only few examples of how the technology is expanding. Throughout the initial and the expansion stages, the ability of a blockchain to store metadata has been exploited in an increasing volume of transactions and with evolving focus of utilization. 

\subsection{Embedding Metadata}
Reviewing historically, and based on the bitcoin blockchain primarily, this ability at first involved creating an Unspent Transaction Output (UTXO) that could never be spent. That was used with a focus on permanently and securely storing information (such as notary data) not directly related to the current transaction. The destination bitcoin address in the locking script of such unspendable UTXO in a Pay-To-Public-Key (P2PK) and Pay-To-Public-Key-Hash (P2PKH) transactions was used as a freeform 20-byte field to store metadata, and the the transaction was recorded on the blockchain. \cite{master} Then, concerns were raised that the unspendable outputs could never be removed from the UTXO database, causing the database to increase forever. In response, Bitcoin Core version 0.9 introduced the RETURN operator, explicitly creating such outputs as provably unspendable and excluded from the UTXO set. \cite{core09} Simultaneously, the allowance for metadata increased from 20 to 80 bytes. Thus, legitimate non-payment data could be stored on the blockchain without increasing the UTXO database. However, concerns were raised that non-payment data stored in OP$\_$RETURN outputs could allow meta-protocols to run permission-free with criminal intent. That led to a large proportion of miners not processing OP$\_$RETURN transactions, and a corresponding proportion of metadata not being recorded on the blockchain. The more recent trend is storing information in the redeem script of pay-to-script-hash (P2SH) transactions. P2SH were standardized with Bitcoin Improvement Proposal (BIP): 16, as a powerful new type of transactions that simplifies the use of complex script. \cite{p2sh} The hash of the redeem script is restricted to 20 bytes, but the script itself and the size of metadata are not restricted. These transactions are spendable, and the full script must be revealed when spending a P2SH UTXO. At this stage, the utilization of embedded metadata is focused on 
variety of applications such as tokenization, blockchain-enforced smart contracts, and related access to secure databases. Some concerns about ransomware remain.

Among the innovative uses of P2SH-embedded metadata are the tokenization of assets (tangible, intangible, divisible, and non-divisible), the blockchain-enforcement of smart contracts, the blockchain-recorded progress through the complex conditionality structure of a smart contract, the efficient blockchain-registered exchange of various tokenized entities underlying  smart contracts, and the blockchain-recorded access links and access privileges to off-chain databases. \cite{contract} Such databases can be Distributed Hash Table (DHT) databases that store smart-contract templates, or conditions for the exchange of and the characteristics of entities underlying smart contracts, or software programs implementing intelligent agents capable of controlling various types of smart contracts. The described utilization of embedded-metadata serves as middleware that supports the development and execution of any specific smart contract.     

In this paper, we identify millions of recorded blockchain transactions embedding metadata and classify them according to meta-protocols they support. That allowed us to observe, from one of its many perspectives, the broader and important question about technology adoption. We further analyze empirically the evolution of embedding metadata, and suggest security steps towards preventing criminal use. Metadata are relevant to any blockchain, and our analysis is based on bitcoin primarily. Bitcoin has long historical data and the largest market-share, but is considered inert to innovation in more recent years.\footnote{Bitcoin has however broken grounds for crypto-currencies, and new currencies, such as bitcoin cash BCH, are actively pursuing innovation.} Our conclusion is that simultaneously with both expanding legitimate utilization of embedded metadata and expanding blockchain functionality, the applied research on improving anonymity and security must continue with protecting against blockchain abuse.

\subsection{Ecosystems}
Bitcoin is the first and has long been the most popular cryptocurrency and blockchain. All\bitcoinA{-transactions} are recorded in the immutable, append only, blockchain data-structure, where the key features and elements include: blocks, transactions stored in the blocks, and inputs, outputs, lock-time included in each transaction according to the transactions' format. The unspent outputs UTXO are monitored by the miners in validating transactions, and each input and output contain scripts -- locking, unlocking, redeem scripts. \cite{Bwiki} Depending on the type of transaction -- P2PK, P2PKH, multisig, P2SH -- the scripts may contain public keys, hashed public keys, multiple public keys, signatures based on private keys, multiple signatures, metadata, hashed metadata, and OP codes. The entire transaction is hashed using SHA-256 and this hash typically serves as a globally unique Transaction ID (TXID). \cite{standard} 

The bitcoin script language \textit{Script} is a Forth-like stack-based execution language. Each transaction is processed by every bitcoin validating node and the node's validation software executes, independently for each of the transaction's inputs, the unlocking script in the input alongside a corresponding UTXO's locking script. A transaction is valid if the cryptographic puzzles in all UTXO referenced by its inputs are solved by the inputs' scripts, i.e. all spending conditions are satisfied. Then these UTXO are removed from the UTXO database but remain permanently recorded on the blockchain. \cite{master} \textit{Script} is a stateless and predictable language, and Bitcoin Core currently includes 174 active \textit{Script} opcodes  (and 15 disabled) including 14 reserved opcodes, of the following types: push-value, flow-control, stack-ops, splice-ops, bit-logic, arithmetic, crypto, locktime, template-matching, and reserved-words. \cite{Bcore}\cite{script} 

The \textit{Script} in the more actively innovating bitcoin-cash BCH-blockchain is introducing further opcodes, by re-designing and re-testing functionalities previously intended (to an extent) by now disabled bitcoin-script opcodes, as well as by introducing new functionalities. \cite{BcashScript} The bitcoin-cash network is undergoing a protocol upgrade in May 2018, supporting on-chain scalability, new transaction signatures, and a new difficulty adjustment algorithm. \cite{BcashUpgrade} The blocksize limit is adaptable, with an increased default of 8MB, and quite larger sizes are being tested on the bitcoin-cash testnet. \cite{BcashTestnet}  A new SigHash reusable signing mechanism ensures replay protection under a chain split, an improved hardware-wallet security, and elimination of the quadratic hashing problem. It provides for users creating transactions with a fork-specific ID, which are invalid on forks lacking support for the mechanism. \cite{sighash}  A new difficulty adjustment algorithm allows miners to migrate from the bitcoin chain as desired and provides protection against hashrate fluctuations. \cite{BcashAdjustment} Multiple independent teams develop bitcoin-cash software, assisted by peer-review workgroups, in contrast to the single-group development of Bitcoin Core. The development of bitcoin cash is decentralized and the ecosystem is dynamic. The focus is on protocol developments and on building Software Development Kits (SDK) that provide for the implementation and support of smart contracts and applications. \cite{sdk}

\section{Improving Privacy and Security}
Extending blockchain functionality and legitimate utilization of embedded metadata demands effective protection against blockchain abuse. The effective protection is supported by active applied research on anonymity and security. 

\subsection{Privacy}
Privacy is a key desirable feature of all public and some private blockchains. Adoption and usage of bitcoin demonstrates early developments in distributed P2P payment systems anonymity engineering, and the privacy levels offered by current bitcoin psuedo-anonymous ledger is not very strong\cite{bonneau}\cite{moser}. Improving this is a major and difficult problem. It is not obvious how to reconcile ledger transparency and the desire for better privacy, and there is no easy quick-fix solution. Some early solutions involve using one public key only once. Using many different keys per user immediately raises the question of key management. In order to avoid the necessity for regular backups of fresh private keys generated at random, deterministic key derivation functions have been introduced. Hierarchical Deterministic (HD) wallets \cite{BIP32} use Elliptic Curve mathematics in order to calculate the public keys without revealing the private keys. HD wallets also allow users to derive various keys in a deterministic way from a single human readable seed. Using several keys there at the same time, however, like joining payments made to several keys belonging to the same user, compromises privacy.\cite{b11} 

In addition to HD wallets, there exist several other methods improving bitcoin anonymity.
\begin{itemize}
\item{\textit{Mixes}: Mixing services or tumblers can be used to improve the anonymity of users by taking their coins and exchanging them with coins of other users, while hiding their identities. These services charge commissions between 1-3 $\%$, and also need to be trusted not to steal users' coins. \cite{bonneau}.}

\item{\textit{CoinSwap}: This is a similar concept to Mixing. If Alice wants to pay Charlie, she can send her coins to Bob instead of Charlie, and then Charlie can send a fresh unrelated coin to Bob. In order to resolve the theft problem, a central authority can manage these swaps. If any of the three misbehaves, the swap may be resolved by using hash-locked transactions that are linkable in the public-ledger.\cite{moser}}

\item{\textit{Fair Exchange}: This method allows the users to hide their identities by exchanging coins. Ideally, the fair exchange requires that either both parties involved in the transition receive each other’s items or none do.\cite{ray}}

\item{\textit{CoinJoin}: In CoinJoin users collaborate and create transactions where inputs of several users are mixed together. The transaction is not valid and will not be accepted by the network until all the signatures are provided \cite{b14}.}
\end{itemize}

Further methods that require attention include stealth addresses and dark wallets:
\paragraph{Stealth Address}
\label{SABasic}
One way to break the linkability in blockchain is to ask a recipient for two destination addresses, and then make two transactions and broadcast them into the network few seconds apart \cite{b11}. This concept has been further improved leading towards Stealth Address (SA) techniques, which are forms of non-interactive key exchange protecting privacy of users receiving payments. The origins of these techniques could be tracked to \cite{CryptoNote20}\cite{bytecoinStealth}. Instead of a public key, the payee advertises a long unique static identifier, which is used by the payer to generate one time address to send money. A stealth address do not appear in the\bitcoinA{-blockchain}; instead, random ephemeral public keys are generated and used. SA addresses use the Diffie-Hellman key-exchange mechanism, which allows the sender and receiver to exchange information and jointly generate some ephemeral public keys. Only lower-level derived keys will appear on the blockchain. 

\begin{figure}[htbp]
\centerline{\includegraphics[width=2.75in,height=1.4in]{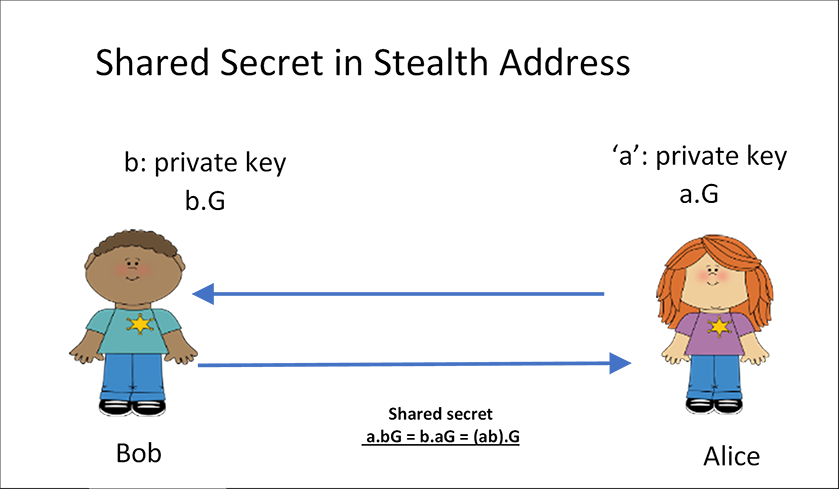}}
\caption{DH key exchange in Stealth Address technique}
\label{secret}
\end{figure}
For example, Bob advertises his (multiple-use) public key on his web page. There exist several variants of SA addressees. A basic method that uses permanent public/private identities of two participants,
Alice and Bob, is adapted from \cite{b12} next. We consider a basic Diffie-Hellman key exchange, as presented in Fig.~\ref{secret}. Let G be the generator point on Elliptic curve, and let 'a' and 'b' be the private keys of the sender (Alice) and the receiver (Bob), correspondingly. Alice and Bob use here their permanent identities that will typically appear on the blockchain (this will change in more advanced SA methods). Alice and Bob publish $a.G$ and $b.G$, correspondingly, and keep their private keys confidential. Alice computes $a.(b.G)$, and Bob equivalently computes $b.(a.G)$. Their shared secret is $S=a.(b.G)=b.(a.G)= a.b.G$, which no one else can compute. Next, Alice sends money to the \textit{transfer address} $E$, and Bob detects the transaction and spends its UTXO. For example:
$$
E=H(S).G
$$
looks like a random\bitcoinA{-addres}. However, Bob knows the corresponding private key as he knows the common secret $S$, and can scan the bitcoin blockchain for $E$ to appear.

This solution is still not quite secure, as Alice also knows the private key and may spend the UTXO before Bob. An improved asymmetric stealth address uses a stronger spending key $e$. \cite{contract} For example,
\begin{center}
$E=H(S).G+b.G$ $ $ $ $ and $ $ $ $ $e=H(S)+b$
\end{center}
Then, the sender Alice can no longer spend the transaction output, and can only compute the public key $E$. This method is still not ideal, as it is static and deterministic. In order to mitigate this, the sender can replace her permanent identity $a$ by a random number $r$, and publish $r.G$ by typically using OP$\_$RETURN in the very next output. In this case, one-time destination key and address are generated. This is not yet the best SA technique, and can be improved further by using 2 public keys, $b.G$ and  $v.G$, where $v.G$ is a view key\footnote{View keys are also used in Monero and other CryptoNote-based currencies, and 
and were first described in \cite{CryptoNote20}.}.
Knowledge of the private part of the view key allows to build read-only wallets that can see transactions (undo anonymity) but cannot spend. Even a more robust stealth address has been proposed that protects against private key compromise, due to thefts, bad random attacks or Spectre/Meltdown type vulnerabilities \cite{b12}. Potential further development can use metadata based on processing various combinations of different partial biometric features, when generating signatures and keys. \cite{bio} Improved approaches to generating hierarchical asymmetric  ephemeral keys and addresses have also been proposed and implemented in Nakasendo, an SDK supporting bitcoin cash applications, as well as applications for any blockchain based on elliptic-curve cryptography \cite{sdk}.

\paragraph{Dark Wallet}
In order to enhance anonymity further, light-weight wallets that use both stealth-address and CoinJoin techniques have been created and termed Dark Wallets (DW). \cite{b20} Stealth addresses are discussed in detail in the previous section. A brief reminder on CoinJoin tells that a transaction of one user is combined with that of a random other user, who is making a payment at around the same time. Dark wallet is currently in its alpha testing state \cite{b21}, as a Chrome extension enabled in developer mode. During this study, it has been working on and off for brief periods of time.

\subsection{Security}
Innovative technologies are subject to abuse, as a series of incidents have demonstrated, including the recent Facebook data abuse by Cambridge Analitica affecting 87 million users.\cite{facebook} Blockchain has also been abused, as the ransomware attack on UK NHS showed last year, affecting many people. \cite{b17} The adoption of bitcoin in ransomware crime is a major event of recent years\cite{b16}. Very recently, the first incident has been reported, as well, of a ransomware accepting bitcoin-cash payments. \cite{BCabuse} In order to protect blockchain expansion into services benefiting the society, it is necessary to address the issue of its abuse. Innovation in terms of security and prevention from ransomware must be an integral part in the development of smart contracts and blockchain-based services. Next, we briefly introduce the key ransomware types and existing defenses. In later sections of this paper, we review them in relation to bitcoin and bitcoin cash, and suggest some solutions. 

\paragraph{Overall Rise of Ransomware}
Ransomware is a class of malware aiming to force users to pay a ransom in order to regain full access to their system \cite{b15}. This terminology covers a wide range of malicious software programs, including CryptoLocker, Locky, Cryptowall, KeyRanger, SamSam, TelsaCrypt, TorrentLocker and others\cite{rAPI}\cite{bonneau} The history of ransomware goes back to 2004, and the early software included screen lockers that were easy to remove or circumvent. Their level of sophistication, however, has been  improving since then. Since 2013, a more harmful type of software has been developed. Though the programs are still called ''lockers'', they are not just lockers but quietly search for specific files and encrypt them, and then ask for ransom in order to decrypt. Only in the last few years, since bitcoin raised in popularity, ransomware has been combined with\bitcoinA{-payments}.\cite{b16}\cite{b17}

\begin{itemize}
\item{\textit{CryptoLocker}: This is a well-known ransomware, since Sep. 2013. CryptoLocker v3 uses Advanced Encryption Standard (AES)-128 in Cipher Block Chaining (CBC) mode\cite{rAPI} and Rivest–Shamir–Adleman (RSA)-2048 for encryption of a header\cite{b19}. AES-128 is a symmetric key algorithm with 128-bit keys, and RSA-2048 is an asymmetric encryption algorithm using 2048-bit keys. This combination makes it most likely impossible to decrypt, without paying the ransom.}

\item{\textit{TorrentLocker Etc.} These are different strains of ransomware that have used AES differently: particularly in Counter (CTR) and CBC modes.\cite{cplockCBCCTR}}

\item{\textit{TeslaCrypt}: This type of malware has been active since 2015, and  provides customer support for the victims. It uses Elliptic Curve cryptography (ECC), an advanced key-derivation scheme, and has an ECC master private key that is later made public.}

\item{\textit{Locky}: Locky is a more recent and more sophisticated ransomware, since 2016. It uses Domain Generation Algorithm (DGA) to prevent blacklisting of domain names, as well as custom encrypted communications. Locky also uses strong (RSA-2048 + AES-128) file encryption, and targets and encrypts
over 160 different file types, including virtual disks, source codes and databases\cite{b18}. Locky has spread in two countries in particular, the United States and France, and uses The-Onion-Router (TOR) hidden servers.\cite{rAPI}} 
\end{itemize}
Further advanced ransomware techniques are discussed in Section \ref{AdvancedRansomOpRetKeyMag}. A recent study by IBM reports 6,000\% of overall increase in ransomware in 2016 compared to 2015, and finds that 70\% of business victim paid the hackers.\cite{b31}  

\paragraph{Ransomware Defenses}
There exist a number of OS-level countermeasures to avoid infection. Such measures include white-listing executables in user data directories\cite{kr2}, avoiding mapping backup drives\cite{cplockCBCCTR}, and disrupting the malware when using the Microsoft Crypto API\cite{rAPI}.

\begin{itemize}
\item{\textit{Data Backups -- False Good Solution}: It may seem that all ransomware is harmless if the data is backed-up on a regular basis and the back-up drives are encrypted. However, the problems go far beyond, and just restoring files or partitions from backups is not the best strategy. This destroys forensic evidence about how the malware propagates and how it operates, and makes the fight again malicious software more difficult. This also leaves our systems wide open and in the same state as before infection: they can be later re-infected by malware through the same channels. For example, a main infection channel for Crypto-Locker was the Gameover Zeus botnet, which had existed earlier.\cite{kr2}}

\item{\textit{Propagation of Ransomware}: Computer security experts must be able to monitor and analyze the infections. It is very useful to know how the ransom gets here in the first place. The problem is that there exist extensive and offensive expertise and experience, which have emerged over years of contrived action against the anti-virus industry. Different types of malware infection propagation and social engineering techniques are exploited to help criminals diffuse their unsolicited encryption payloads.}
\end{itemize}

\section{Use and Abuse of Blockchain Transactions: Empirical Analysis}

\subsection{Methodology and Results}
The length of OP$\_$RETURN script is currently 80 bytes, where the first two bytes always are hex 6a, followed by two bytes indicating the length in hex of the metadata-record that starts from byte number 5. With its protocol upgrade from version 1.0 to 1.1, on May 15, 2018, the OP$\_$RETURN relay size will increase to 223 bytes, only on the bitcoin-cash blockchain.\cite{BcashUpgrade} Information about the two blockchains has been updated, corrected and analyzed in this study, by the contributing authors, based primarily on the OpReturnTool from\cite{b23}.  

The APIs of blockchain.info and coinsecrets.org are queried by this software. Blockchain.info is used only to get the latest block number. Then coinsecrets.org, a dedicated API for OP$\_$RETURN transactions, is queried to extract their Time stamp, Transaction ID, hash and ASCII code. The incoming data are recorded into a text file and exported to Excel\cite{tooba}. Several methods were used, in order to identify the evolution of stealth-address techniques. We have performed experimental transactions, and further analysis on the observed patterns, to identify potential DW  transactions and patterns. Among those that could not be related to a known protocol, there are transactions potentially related to criminal activities. 

To identify the transactions from another wallet implementing stealth address, such as SX\cite{b28},
its documentation has been consulted and observed patterns inside transactions are matched with our dataset. Overall, 22 protocols are identified and the rest of the OP$\_$RETURN transactions are marked as “unattributed”. The prefix and ASCII for all the protocols are analyzed, and based on the analysis further three protocols have been identified: YEJ, BITCC, and Counterparty. However, their share inside the dataset is quite slim ($\approx 0\%$), and only Counterparty has been included for further analysis.
\begin{figure}[htbp]
\centering
\includegraphics[width=3in,height=1.5in]{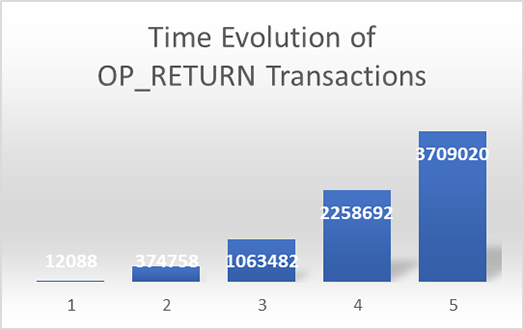}
\caption{Time evolution of OP$\_$RETURN transactions}
\label{evolution}
\end{figure}

The first OP$\_$RETURN transaction, identified in our study, appears on Mar. 29,  2013. This corrects \cite{moser}, where the first such transaction is identified as appearing a year later. Thus, our dataset consists of data since Mar. 29, 2013 (block \#228596) to Jul. 6, 2017 (block \#474451). In 2013, only 430 OP$\_$RETURN transactions are found, and all of them are in the unattributed category. From 2014 to 2017, we observe a significant increase in the volume of such transactions, reaching over 2 million per year in 2017. (see Fig. \ref{evolution}) A detailed examination shows that $\approx 51\%$ of these transactions correspond to the 22 known protocols, explained briefly in Table~\ref{protocols}. About $49\%$ of the transactions remain unattributed.
\begin{table}[htbp]
\caption{Protocols Using op$\_$return Opcode}
\begin{center}
\begin{tabular}{|c|c|c|}
\hline
\textbf{Protocol}& \textbf{Contribution(\%)}& \textbf{Usage}\\
\hline
Unattributed& 49\%& Not identified\\
\hline
Blockstore& 8.5\%& Key value store\\
\hline
Factom& 4.14\%& Notary/Doc\\
\hline
Omni Layer& 10.3\%& Assets\\
\hline
Blocksign& 0.06\%& Notary/Doc\\
\hline
Colu& 10.11\%& Assets\\
\hline
Stampery& 2.60\%& Notary/Doc\\
\hline
Eternity wall& 0.16\%& Any Messages\\
\hline
Bitproof	& 0.03\%& Notary/Doc\\
\hline
Open Assets& 8.09\%& Assets\\
\hline
Ascribe& 2\%& Digital Arts\\
\hline
Monegraph& 2.7\%& Digital Arts\\
\hline
Coinspark& 1.1\%& Assets\\
\hline
Proof of Existence& 0.22\%& Notary/Doc\\
\hline
Original My& $<$0.01\%& Notary/Doc\\
\hline
Open Provenance& $<$0.01\%& Proof of ownership\\
\hline
Remembr& $<$0.01\%& Notary/Doc\\
\hline
Crypto copyright& $<$0.01\%& Notary/Doc\\
\hline
LaPreuve& $<$0.01\%& Notary/Doc\\
\hline
ProveBit& $<$0.01\%& Notary/Doc\\
\hline
Blockchain Notary& $<$0.01\%& Notary/Doc\\
\hline
Counterparty& $<$0.01\%& Assets\\
\hline
Stampd& $<$0.01\%& Notary/Doc\\
\hline
\end{tabular}
\label{protocols}
\end{center}
\end{table}

\begin{figure}[htbp]
\centering
\includegraphics[width=2.9in,height=1.7in]{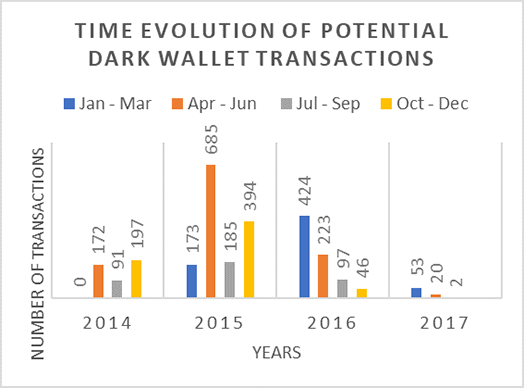}
\caption{Time evolution of potential DW transactions}
\label{wallet}
\end{figure}

The experimental dark-wallet transactions are identified in the unattributed section of the dataset, and have the prefix 6a-26-06, where 6a is the opcode for OP$\_$RETURN, 26 is hex of the length of metadata that follows, and 06 distinguishes dark-wallet transactions from other protocols. The whole dataset is scanned and $\approx 2762$ transactions are found with this pre-fix. The time evolution of transactions is illustrated with Fig.~\ref{wallet}.

\subsection{Analysis and Discussion}
Linkability of transactions affects their anonymity, and several techniques have been adapted in the\bitcoinA{-systems} to address that issue. Approaches such as CoinJoin, Fair Exchange and CoinSwap, typically need a third-party involvement to achieve anonymity, and the honesty of the third party is not guaranteed. The latest stealth-address techniques seem currently effective. There, the receiver advertises its static, unique identifier and the sender generates a one-time key. There is no apparent way that a blockchain observer can relate transactions to the same payee. Stealth-address approaches are introduced to bitcoin and bitcoin cash, and have been used in monero and vertcoin. Stealth-address, dark-wallet and SX transactions appear as ‘unable-to-decode’ by block explorers.

\begin{figure}[htbp]
\centering
\includegraphics[width=3in,height=1.5in]{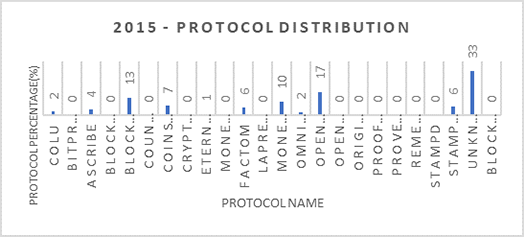}
\includegraphics[width=3in,height=1.5in]{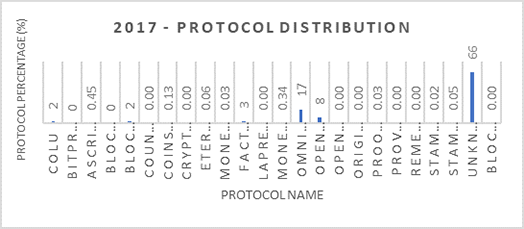}
\caption{The share of unattributed protocols is increasing.}
\label{distribution}
\end{figure}

We have identified some DW and SX transactions among transactions unattributed to known protocols, in the OP$\_$RETURN database we have extracted from the blockchain. It is noted that a smaller number of SX-related transactions are identified, as that protocol is less user-friendly than DW. A large number of transactions remain unattributed, and part of the issue is that meta-protocols are not required to coordinate and register unique identifiers. Therefore, many legitimate protocols don't use distinctive pre-fixes\cite{b24} that help decode/classify OP$\_$RETURN transactions they produce, and those transactions remain unattributed. This reason is even more valid now and in foreseeable future, as the anticipated proliferation of blockchain technology is through smart contracts that benefit users rather than ransomware them, and brings positive rather than destructive effect on society. Smart contracts are implemented and executed through embedding metadata in OP$\_$RETURN and P2SH transactions, and therefore the number of transaction with embedded metadata will continue to rise. It is also valid to anticipate that criminals will exploit the new functionalities. We address both these issues in Section ~\ref{RansomVsBitcoin} next. 

\section{Ransomware, Key Management and Metadata}
\label{RansomVsBitcoin}
\subsection{Ransom Payments}
Ransomware has always existed, but it has been associated with substantial risks to receive ransom payments without detection. With the increased popularity of Bitcoin in the last few years, criminals have started abusing the technology, in order to avoid detection.\cite{b30}. Receiving one or multiple ransom payments in\bitcoinA{} allows very good initial anonymity.\cite{b16}. A new unique\bitcoinA{-address} is created to receive payment from each victim. As long as the coins are not yet spent, there is no way to track who has received the ransom, and it can be spent in the future.\cite{b15} Once spending starts, the linkability of transactions is weakening anonymity and some transactions could be traced.\cite{b7} However, criminals can use various proxies, and carefully move and mix money in arbitrary ways for a long time, in order to diminish their chances of being traced. 

\begin{figure}[htbp]
\centering
\includegraphics[width=3.6in,height=2in]{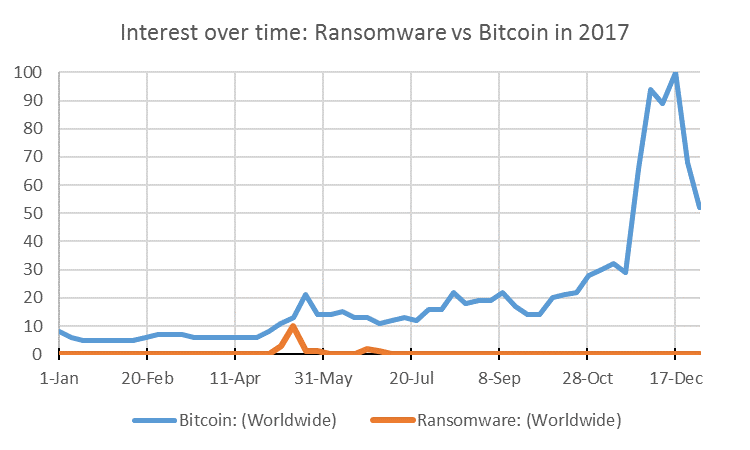}
\captionsetup{justification=centering,margin=0.2cm}
\caption{Google search interests: Ransomware vs Bitcoin January--December 2017}
\label{google}
\end{figure}

Many companies and individuals will and do pay ransomware to get their data back, though advised the contrary by the authorities. Some individuals bought\bitcoinA{} for the first time when became victims of ransomware, and some companies buy\bitcoinA{} in advance to be able to pay in case of an attack.\cite{b30} This affects the image of the technology, as illustrated by Google Trends during the attack on NHS last May. Fig.~\ref{google} presents the online interest in bitcoin and ransomware for the first two quarters of last year, and identifies they both spiked in May. \cite{b33} The image affects the development and adoption of innovative and legitimate blockchain-based services. 
The technology is still moving forward, however, as Fig.~\ref{google}, because the positive potential has been recognized. The target now is decoupling and disassociation from ransomware.

\subsection{Public Key Generation and Diversification}
\label{AdvancedRansomOpRetKeyMag}
A variant of the Curve-Tor-Bitcoin (CTB)-Locker, targeting web servers, generates a unique\bitcoinA{-address} for every infection. Once the ransom is received, hackers produce a transaction using OP$\_$RETURN and embedding a decryption key inside.\cite{b34}
Other ransomware, such as the Locky payment system, rely on the anonymity features of TOR. It uses TOR hidden servers that remain operational years after the infection. The server software, presumably operating without human intervention, has automatically adjusted to\bitcoinA{-price} over the years: initially asking for\bitcoinA{2}, and recently for much smaller amounts. A Locky decryptor tool, made available to the victims on the same TOR website when they connect again after paying, is received on the blockchain\cite{b18}. One of the onion servers used with Locky is \url{twbers4hmi6dx65f.onion}, and we have observed that after payment they contain malware such as Variant.Zusy.185950. The Locky payments also seem to be automatically aggregated into larger pots of \bitcoinA{50 or 100}. An address associated with Locky ransom is \url{1Cjqt4C17sXYrWkrRyPr73RjrjZu1fuHMV}, though it is not clear if the address belongs to an exchange, a mixing service, or is still criminally controlled. If we look from this address backwards, the blockchain allows identifying victims of Locky who have paid ransom in standardized amounts of\bitcoinA{1 or 0.25}.

\subsection{Future Risks and Mitigation}
The combination of all topics we study above present a major risk: using stealth addresses, CoinJoin, TOR, and possibly smart contracts, in order to further automate and streamline the process of ransom threat and payment. We are not there yet but in future, criminals can hide their identity behind the increasing number of active blockchain participants, while shuffling money around.
\begin{figure}[htbp]
\centering
\includegraphics[width=3.5in,height=1.8in]{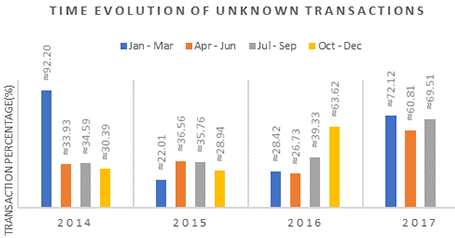}
\caption{Share of unattributed protocols.}
\label{unkonown}
\end{figure}
Dark wallet has been a prominent\bitcoinA{-wallet}\cite{b21} focusing on anonymity. It implements stealth-address with CoinJoin techniques. Yet DW deployment has raised concerns about its potential abuse.\cite{b20} This study has analyzed DW transactions and found that they cannot be associated with known protocols. The number of DW transactions is still low: the wallet is in its alpha state and has not recently been operating properly.\cite{b21} It can be expected, however, that a more dangerous DW-successor will be inevitably created.\cite{bytecoinStealth} We propose that companies using blockchain to implement complex protocols, should provide means of audit capable of distinguishing legitimate traffic they produce from potentially criminal activity.

Though bitcoin and bitcoin cash are studied here, all blockchains are imminently subject to abuse, and fall within the wider range of FinTech and SocTech technology targeted with criminal intent. Bitcoin has existed longer and had the largest market-share among blockchains. On the one hand, it has traversed through challenges to bring visibility to the technology and recognition of its potential, and opportunities to and drive for emerging competition. On the other hand, this position has made bitcoin most targeted by criminal activity: the wider and more active is a blockchain network, the better it can be both used and abused. Empirical history has also shown that the more a cryptocurrency is used, the more valuable it is. Further blockchains are getting momentum and raising their market-share, and are imminently attracting criminal intent. The other case-study in this paper, bitcoin cash, is representative of recently set up but high-momentum blockchains, and has risen to 4th market-share. Only one ransomware has been reported using bitcoin cash, seven months after its launch. This contributes to the observation that relative maturity is a precondition for targeted abuse of a blockchain.

With this paper, we raise the requirement that security and robustness against ransomware and other abuse must be of equal priority in innovation as are functionalities directly impacting market-share. Another factor is the rate of innovation itself. For example, blockchain cash is supported by several rather than one development team, and releasing a new protocol this May that introduces new op-codes, increases blocksize, and provides a new algorithm stepping towards blockchains' interoperability. The need for innovation has been a main reason for the bitcoin cash fork. The development teams have started releasing SDKs to support the blockchain community in implementing higher-level apps, as well. Multiple teams and continuous innovation contribute to resistance to abuse. We have to note that a large part of identified transactions without attributed protocols are associated with legitimate meta-protocols, but there is no registry of meta-protocols to serve as a reference for identification. The use of P2SH transactions and the development of smart contracts are trends for a foreseeable future, and will expand significantly the range of meta-protocols. Therefore, we suggest that an off-chain DHT registry of meta-protocols is set up, and a corresponding unique indicator/identifier for each type of protocol is embedded in transactions that use/implement it. The DHT repository will have secure selective access, so that the meta-protocols are not compromised but allow corresponding level of audit. We anticipate that with the proliferation of smart contracts, the legitimate types of meta-protocols will greatly outnumber malicious ones. However, the system will still be vulnerable to large-scale outlier cyber-attacks, if decisive, expert, priority solutions are not developed, maintained and updated. If anticipation is that criminals will use smart contracts, then monitoring smart contracts can be developed to identify, protect against, and prevent their activity. 

\section{Conclusion}
This paper addresses blockchain adoption from the perspective of the evolution and utilization of metadata, and therefore, from the perspective of the evolution of protocols that are implemented and executed through embedding additional information in blockchains. We identify, analyze and discuss reasons for the role of metadata, including empirical analysis; and suggest approaches towards protecting the expanding blockchain functionality against criminal abuse and ransomware. Challenges, in terms of exploiting/improving anonymity and prioritizing/raising security, are clearly stated, and intended and unintended consequences are addressed. We review key characteristics of ransomware and of expert approaches protecting against it. The paper suggests that blockchains should provide some level of transparency of what they are used for. We need to improve the auditability of blockchain transactions and smart-contract protocols. This can be facilitated for example by DHT databases of protocols, with secure and selective authorization or audit access. Another major lesson learned from our research on blockchain technology adoption and on metadata usage in blockchains, is to see that new technology must innovate and change all the time in order to near its full potential and that protocols and usage of blockchains must evolve with time. We need to continue to improve privacy of blockchains, while mitigating and monitoring their problematic or maybe criminal usage.

\end{document}